\documentclass[a4paper,11pt]{article}
\pdfoutput=1 

\usepackage{art}

\title{\boldmath Developing high-performance DIRC detector for the Future Electron Ion Collider Experiment}

\author[a,1]{G.~Kalicy,\note{Corresponding author.}}

\affiliation[a]{The Catholic University of America, Washington, USA,}
\emailAdd{kalicy@cua.edu}

\abstract{
  The Electron-Ion Collider (EIC) will be the next frontier project of nuclear physics in the United States. It is planned to be built in the Brookhaven National Laboratory (BNL) in close collaboration with Jefferson Lab. One of the key requirement for the EIC central detector is excellent particle identification (PID). A detector using the Detection of Internally Reflected Cherenkov light (DIRC) principle, with a radial size of only  7-8~cm, is a very attractive solution for identification of the hadrons in the final state. The R$\&$D program performed by the EIC PID collaboration (eRD14) is focused on designing a high-performance DIRC (hpDIRC) detector that would extend the momentum coverage well beyond the state-of-the-art 3 standard deviations or more separation of $\pi/K$ up to 6~GeV/$c$, $e/\pi$ up to 1.8~GeV/$c$, and $p/K$ up to 10~GeV/$c$. Key components of the hpDIRC detector are a 3-layer compound lens and small pixel-size photo-sensors. This article describes the status of the high-performance DIRC R$\&$D for the EIC detector, with a focus on efforts towards developing and validating the radiation hard 3-layer lens.
}


\collaboration[c]{on behalf of for the EIC PID Collaboration for an integrated program for Particle Identification (PID) at a future Electron-Ion Collider}

\proceeding{INSTR20 - The International Conference "Instrumentation for Colliding Beam Physics" \\
  Novosibirsk, Russia}

\begin{document}
\maketitle
\flushbottom

\section{Introduction}
\label{sec:intro}
The Electron Ion Collider (EIC) is the future for the nuclear physics research in the United States. It will be the world's first collider with polarized electron and light ion beams and capable of heavier, unpolarized ion beams up to uranium. The broad and ambitious physics program of the EIC is described in the White Paper~\cite{WhitePaper}.
Excellent hadronic PID in the central detector over a large range of angles and momenta is essential for the study of exclusive and semi-inclusive processes, which allow to image the 3D structure of the nucleon. It is also important for heavy flavor physics.

\begin{figure}[tbh]
\includegraphics[width=0.95\textwidth]{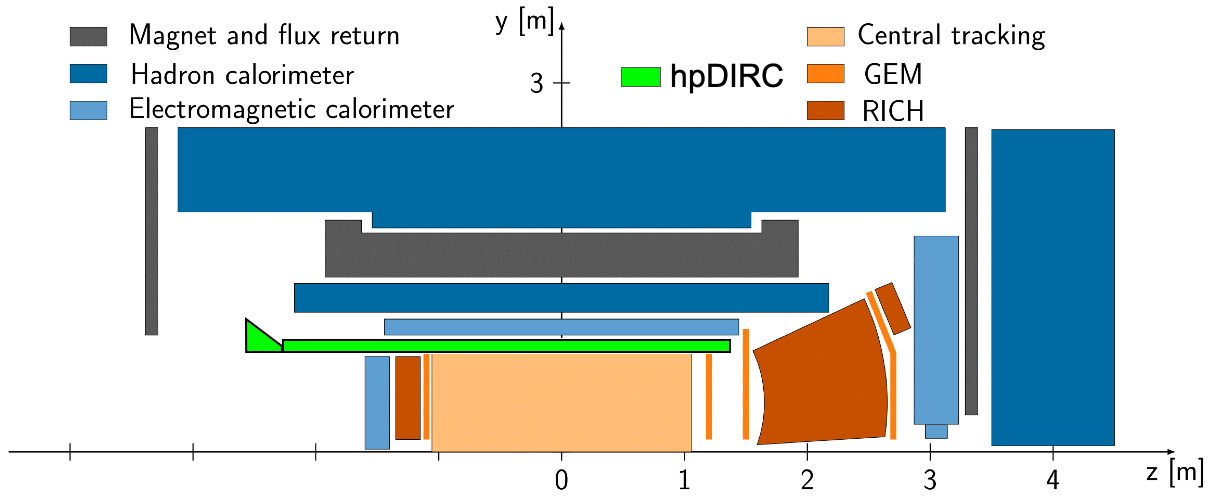}
\caption{Plan view of the layout of the Brookhaven version of a full-acceptance EIC detector based on sPHENIX. The hpDIRC detector is shown in green in the central barrel section. }
\label{fig:EIC}
\end{figure}

The ambitious physics program of the EIC detector requires ability to identify hadrons in the final state, and tag the flavor of the struck quark in semi-inclusive deep inelastic scattering (DIS) to provide information about the transverse momentum distributions.

Three alternative concepts of EIC central detectors are being developed at BNL and Jefferson Laboratory (JLab) but they all assume detector based on the Detection of Internally Reflected Cherenkov light (DIRC) concept in the barrel region. As an example, the BNL EIC full-acceptance detector based on sPHENIX, with incorporated DIRC detector, is shown in Fig.~\ref{fig:EIC}. Originally, they had slightly different layouts of the hadron ID systems, but the configurations have converged. The PID consortium is developing an integrated solution that would be suitable for the EIC physics requirements. The selection and arrangement of the sub-detectors is optimized to detect and identify the complete final state of a nuclear reaction, including all partonic and nuclear fragments.  

\begin{figure}[htb]
\centerline{%
\includegraphics[width=0.95\textwidth]{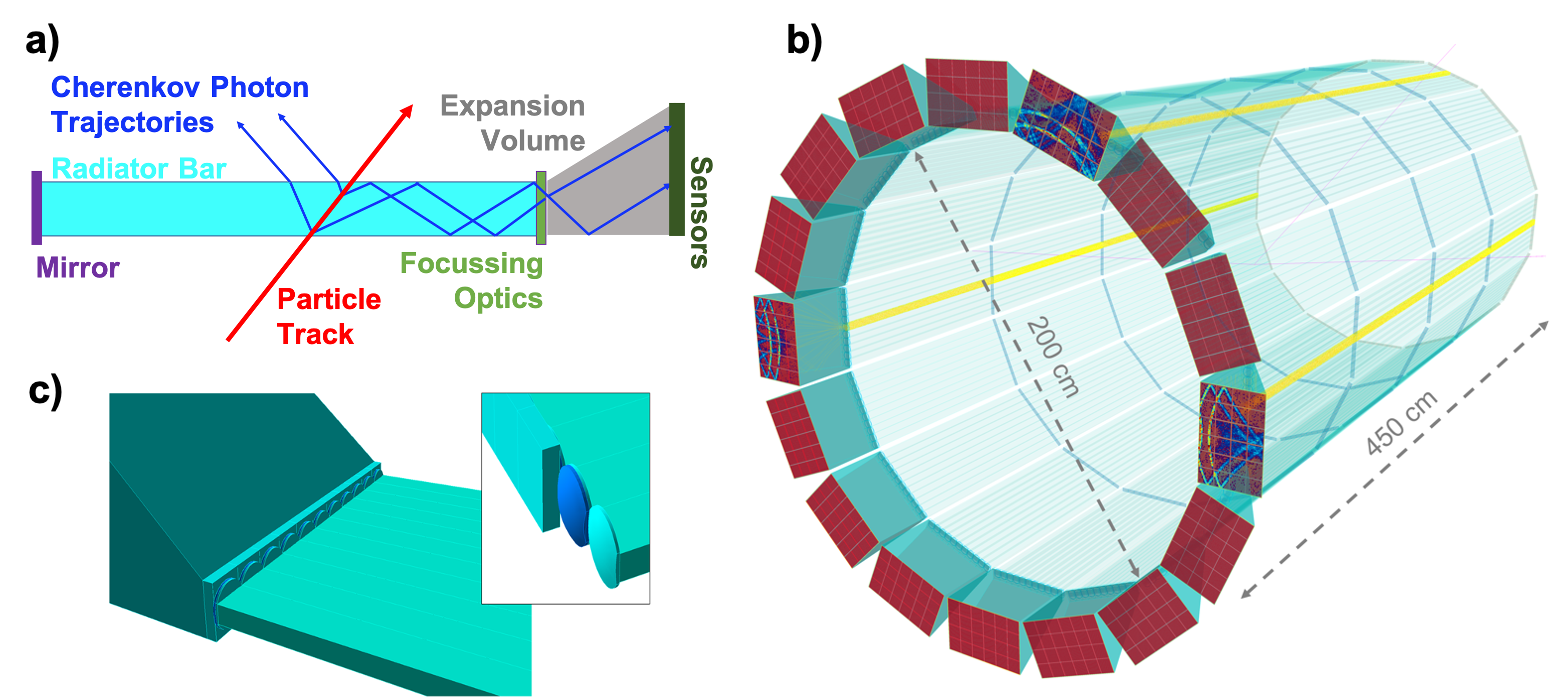}
}%
\caption{ a) Drawing of the DIRC principle. b) Geant4 geometry for the simulation of the hpDIRC accumulated Cherenkov photon hit pattern for charged kaons. c) 3D visualization of the fused silica prism expansion volume, a row of spherical 3-layer lenses and the radiator bars. The insert shows the individual lenses and layers of the spherical lens system.
 }
\label{fig:DIRCprinc}
\end{figure}

The detector based on the DIRC principle, shown in Fig.~\ref{fig:DIRCprinc}a, meets all the requirements to identify hadrons with the lowest energies in the barrel region. Because it will be surrounded by the electromagnetic calorimeter the amount of material before the calorimeter measured in Radiation lengths must be minimized. It also has to be able to operate in the fringe field of the solenoid, a non-uniform magnetic field of a magnitude of 1.5~T or higher. 
The DIRC detector is a special kind of RICH counter using solid rectangular-shaped radiators made of synthetic fused silica that are utilized also to guide Cherenkov photons to the readout section, placed on the side where the photons are recorded by an array of pixelated photon sensors. Thanks to the excellent optical finish of the optics the emission angle of Cherenkov photons, in respect to the particle track, is maintained during the photon transport via the total internal reflection and can be reconstructed from the measured position of the photon on the detector surface and the arrival time of each photon.

Different applications of DIRC technology are used in a number of new particle detectors but they all originate from the first DIRC detector used in the BaBar experiment~\cite{BaBar}, successfully operated at SLAC for over 8 years, providing excellent hadronic PID. The PANDA experiment at the FAIR facility, now entering the construction phase, designed a lens-based focusing DIRC detector~\cite{PANDAdirc} for the barrel region of the target spectrometer. The PANDA Barrel DIRC detector introduced significantly more compact design than the BaBar DIRC but will maintain its great performance by using focusing lenses and faster electronics. 

The EIC DIRC concept is inspired and takes advantage of the PANDA Barrel DIRC detector, but aims to almost double the momentum coverage, well beyond the current state-of-the-art. This paper will discuss few aspects of the R$\&$D program with the focus on developing the radiation hard lens.

\section{High-performance DIRC Detector}\label{sec:Design}

To separate pions from kaons at the 6~GeV/$c$ momentum, with a separation power of at least 3 standard deviations required for the EIC detector, the compact high-performance DIRC (hpDIRC) detector is being developed. It is currently considered by all EIC central detector concepts described in Ref.~\cite{HandBook}. The geometrical details of the DIRC differs slightly for each central detector concept but it does not influence the final performance.

The initial design of the hpDIRC detector was implemented in a detailed Geant4 simulation and is shown in Fig.~\ref{fig:DIRCprinc}b, together with three example Cherenkov ring patterns. 
It is divided into optically isolated sectors that comprise a bar box and a readout box, surrounding the beam line in a 16-sided polygonal barrel at a radius of 1~m.
Each bar box includes a set of eleven radiator bars, made of synthetic fused silica bars, each 4200~mm long, with a cross section of 17~mm $\times$ 32.7~mm. The bars are placed side-by-side, separated by small air gaps, in a light-tight container. 
Mirrors are attached to one end of each bar to reflect Cherenkov photons towards the readout end, where they exit the bar and are focused by a 3-layer spherical lens on the back surface of the prism that serves as an expansion volume. A detailed view of the readout end of the bar box is shown on  Fig.~\ref{fig:DIRCprinc}c. The prism expansion volume is made of synthetic fused silica, has a $32^{\circ}$ opening angle and dimensions of 237~mm $\times$ 360~mm $\times$ 300~mm. The detector plane of each prism is covered by 3~mm $\times$ 3~mm pixels, selected as a compromise between cost and performance for the hpDIRC design based on the Geant4 study~\cite{EICDIRC}, to record the position and arrival time of the Cherenkov photons. Several options for the sensors are still being considered and developed. Sensor options under consideration are MCP-PMTs and SiPM. For the more realistic simulation of the hpDIRC performance, parameters like quantum efficiency of Photonis Planacon XP85012~\cite{Photonis} MCP-PMTs are used. The hpDIRC design can be easily scaled to smaller diameters and bar lengths. Other design options are also being studied, like for example replacing eleven narrow radiator bars in the barbox with one wide plate. It would significantly lower the cost of the final detector but presents additional challenges to the reconstruction approach. 

\begin{figure}[htb]
\centerline{%
\includegraphics[width=0.95\textwidth]{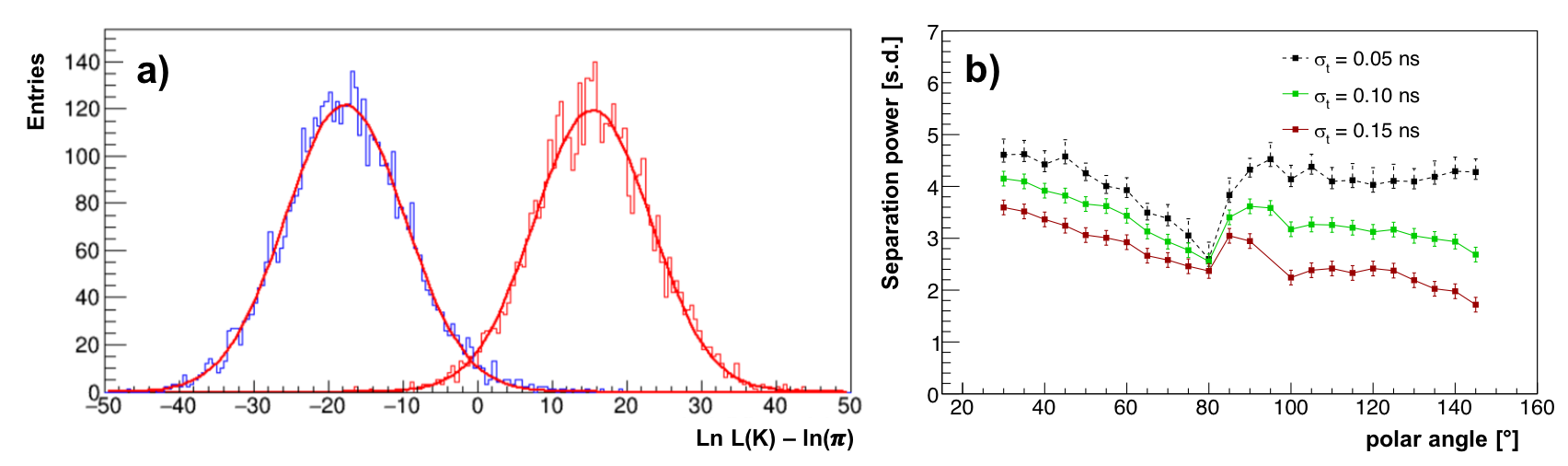}
}%
\caption{ Simulated time-based imaging PID performance of the High-Performance DIRC. a) Log-likelihood difference for kaon and pion hypotheses for a sample of 6 GeV/c pions and kaons at 30$^{\circ}$ polar angle. The $\pi/K$ separation power extracted from the Gaussian fits is 4.2~s.d. b) $\pi/K$ separation power as a function of the polar angle. Different colors represent the assumption of different time resolutions. Plots previously published in~\cite{EICDIRC}.}
\label{fig:simperf}
\end{figure}

The different design options of the hpDIRC are evaluated using a specially developed "time-based imaging" reconstruction method based on the approach used by the Belle II time-of-propagation (TOP) counter~\cite{TOP}.  A large number of simulated tracks with the defined momentum, location, and charge of the particle are used to generate a probability density functions (PDF). All possible Cherenkov photon hits and their arrival times are saved for each pixel on detector plane. For a given track the observed photon arrival time for each hit pixel can be compared to the histogram array to calculate the time-based likelihood for the photons to originate from a given particle hypothesis. In current simulation studies the optimistic, realistic, and pessimistic scenarios of the per-photon timing precision of the sensor and readout electronics were studied, respectively 50, 100, and 150~ps.

Figure~\ref{fig:simperf}a shows an example of the log-likelihood difference for kaon and pion hypotheses for a sample of simulated 6~GeV/$c$ pions and kaons at 30$^{\circ}$ polar angle and the photon timing precision of the sensor plus electronics was set to 100~ps. The $\pi/K$ separation, derived from the difference of the two mean values of the fitted Gaussians divided by the average width, corresponds to 4.2 standard deviations in this case. Figure~\ref{fig:simperf}b shows $\pi/K$ separation as a function of polar angle for the full phase space assuming descried above three different time resolution scenarios. Precision matters more for shorter photon paths (polar angles above 100$^{\circ}$).  The realistic scenario with the 100~ps timing precision provides a required 3~s.d. $\pi/K$ separation for the full phase space. 

\section{Development of the 3-layer Lens}

\begin{figure}[htb]
\centerline{%
\includegraphics[width=.95\textwidth]{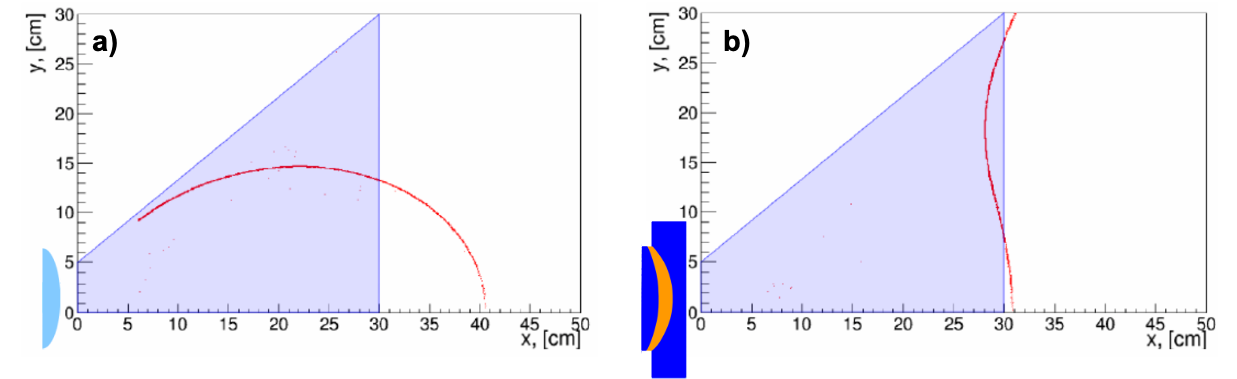}
}%
\caption{ Simulated shape of the focal plane (red points) overlaid on the prism shape and drawings of corresponding lens: a) air gap lens b) spherical 3-layer lens.}
\label{fig:LensPlane}
\end{figure}

A crucial component of the hpDIRC detector, that allows it to reach the desired performance, is the novel 3-layer lens. A standard focusing lens with the air gap at the transition from the lens to the expansion volume would cause a dramatic photon loss. A compound lens made of the combination of the synthetic fused silica and high-refractive index material in 3-layer lens makes the difference of refractive index smaller, minimizing the photon loss. Figure~\ref{fig:LensPlane} shows second challenge that the 3-layer lens is resolving. The DIRC imaging plane is located on the flat back surface of the prism. The focal plane of a standard lens with a single refracting surface has a parabolic shape described by the Petzval field curvature~\cite{Petzwal}. The curved focal plane makes major part of the Cherenkov ring out of focus, significantly deteriorating the performance. The complex multi-layer spherical compound lens allows to shape the focal plane to better follow the detector plane by introducing a combination of focusing and defocussing lens system.

\section{Mapping of the Focal Plane}

\begin{figure}[htb]
\centerline{%
\includegraphics[width=.95\textwidth]{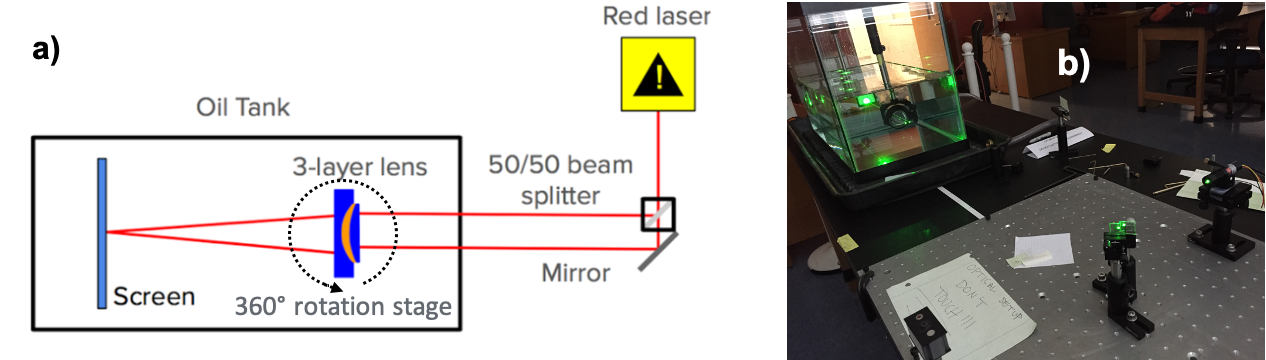}
}%
\caption{ Schematic diagram (a) and the photo (b) of the laser setup to map out the focal plane of the 3-layer lens as a function of rotation angle of the lens.}
\label{fig:LensSetup2}
\end{figure}

 Several 3-layer lens prototypes were designed and build by different vendors. Their focusing properties have to be validated on the test bench to make sure the desired shape of focal plane was achieved. Figure~\ref{fig:LensSetup2}a shows the schematic of the laser setup designed and built in the Old Dominion University to map the shape of the focal plane of the lens prototypes. To simulate the lens being placed between the synthetic fused silica bar and the prism in the mapping setup the lens, mounted on the rotation stage, are placed inside a glass container filled with mineral oil~\cite{oil}, with a refractive index very close to fused silica. The focal length is determined by finding the intersection point of the two parallel laser beams. Two lasers were used in the measurements red (670nm) and green (532nm). To compare spherical and cylindrical lens designs a special 3D-printed holder is used that allows rotation of the lens in two planes.
 
 \begin{figure}[htb]
\centerline{%
\includegraphics[width=.95\textwidth]{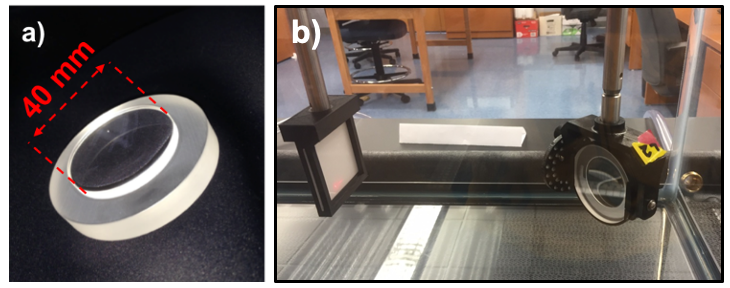}
}%
\caption{ Photo of the spherical 3-layer lens prototype with marked size (a) and in the laser setup (b).}
\label{fig:spherical}
\end{figure}

\begin{figure}[htb]
\centerline{%
\includegraphics[width=.95\textwidth]{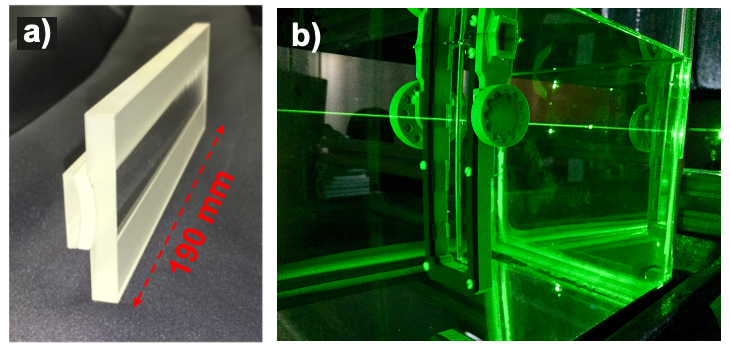}
}%
\caption{ Photo of the cylindrical 3-layer lens prototype with marked size (a) and in the laser setup (b).}
\label{fig:cylindrical}
\end{figure}

Two prototype lenses with middle layer made of the Lanthanum crown glass, with a refractive index of  1.786 at 380nm, were build and measured in the laser setup.  Figure~\ref{fig:spherical} shows photos of the round spherical lens prototype that can accommodate a single narrow radiator bar. Figure~\ref{fig:cylindrical} shows photos of the cylindrical lens prototype that can be used also in the wide plate radiator geometry of the DIRC detector.

\begin{figure}[htb]
\centerline{%
\includegraphics[width=.95\textwidth]{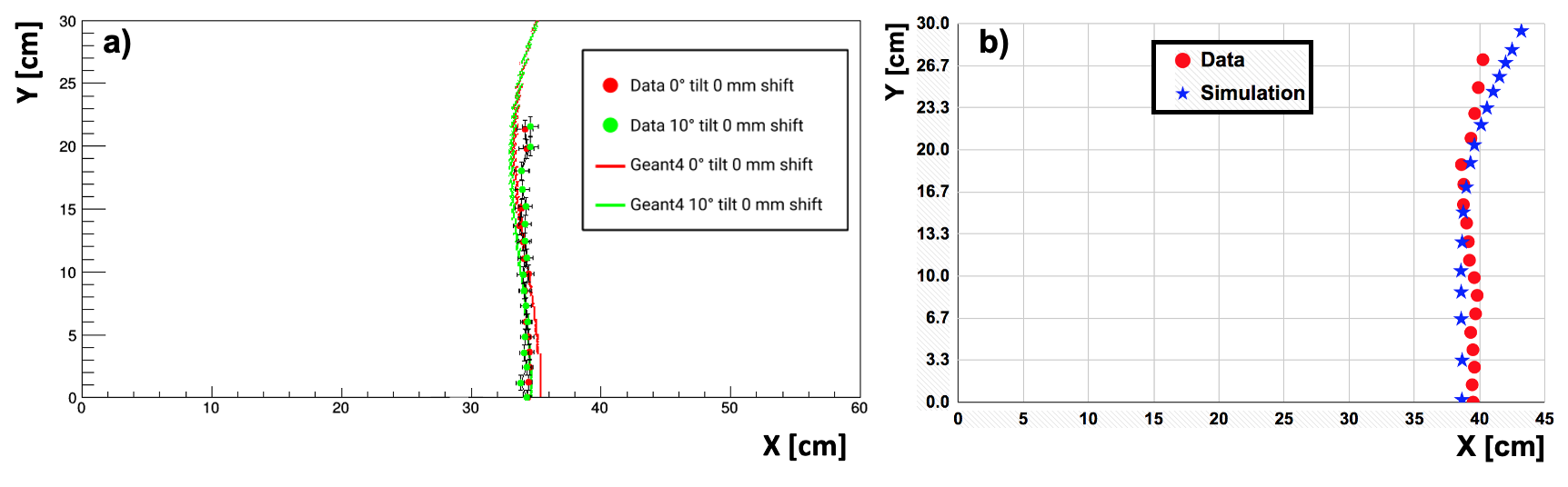}
}%
\caption{ Measured shape of the focal plane compared to simulation for spherical (a) and cylindrical (b) 3-layer lens prototypes. The x coordinate corresponds to the focal length and the y coordinate to the focal point location perpendicular to the incident beams, calculated for different lens rotation polar angles}
\label{fig:FocalMap}
\end{figure}

The results of the focal plane measurement are compared to the predictions from the GEANT4 simulation in Fig.~\ref{fig:FocalMap}, where both laser beams were centered on the lens, separated by 2.5~mm. During these measurements, the lens was set perpendicularly to the laser beam plane, the lens was rotated from 0$^{\circ}$ to 35$^{\circ}$. The measured shape of the focal plane is in very good agreement with the simulation. Both lenses are producing the desired flat shape of the focal plane. Additional measurements for other rotation angles and off-center laser beam positions are planned.

\section{Radiation Hardness of the Lens Materials}

The outer layers of the 3-layer lens prototypes are made of synthetic fused silica. Synthetic fused silica, which is used for most of the optical components in all DIRC systems, was already extensively tested for the BaBar and PANDA DIRC counters and proved to be radiation hard~\cite{BaBarRad}. The 3-layer spherical lens designed for the PANDA Barrel DIRC will use lanthanum crown glass, which was not known to be radiation hard, but the modest dose of only 4 Gy during 10 years of operation~\cite{Carsten} is significantly lower than predicted for the EIC. Due to higher doses predicted for EIC, the validation of all new materials is critical for the hpDIRC development.  

\begin{figure}[htb]
\centerline{%
\includegraphics[width=.95\textwidth]{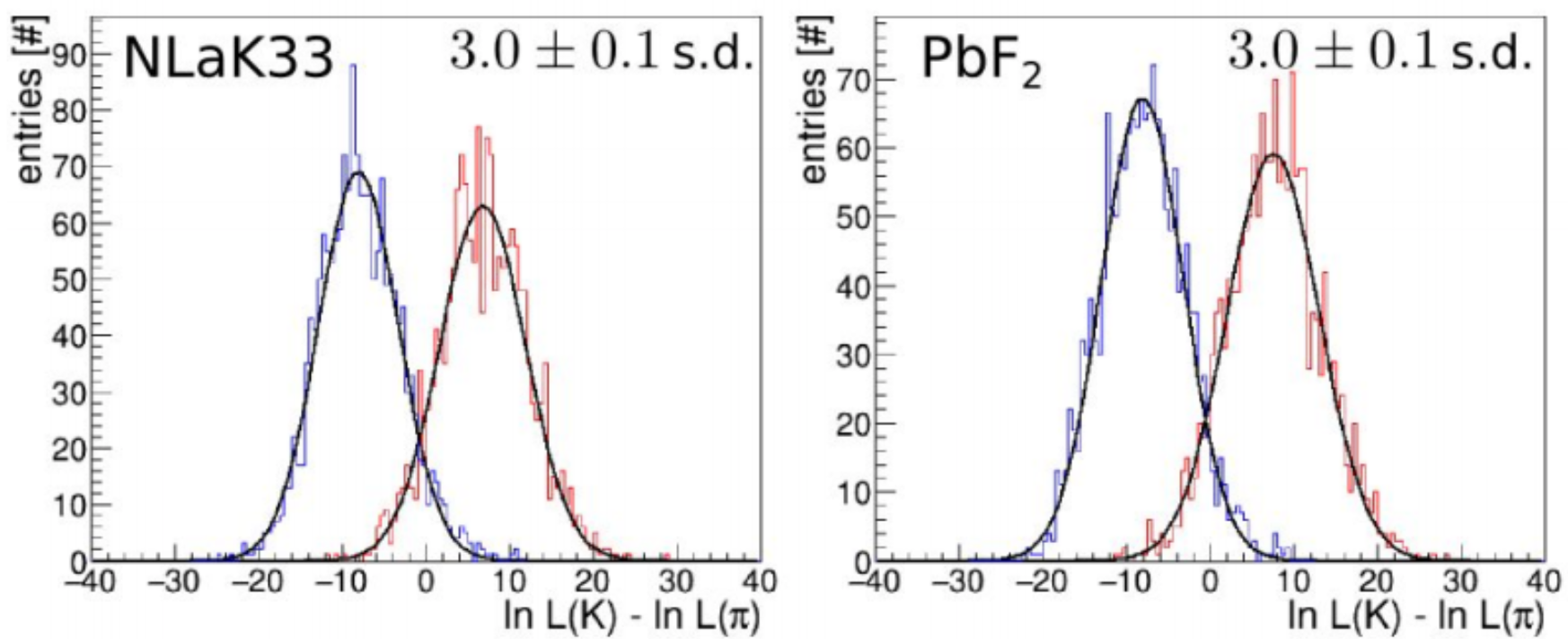}
}%
\caption{Comparison of the simulated hpDIRC performance with two possible materials for the middle layer of the 3-layer lens: NLaK33 (a) and PbF$_{2}$ (b). $\pi$/K separation for charged pions and kaons with 6~GeV/c momentum and 30$^{\circ}$ polar angle, assuming a tracking resolution of 0.5 mrad. }
\label{fig:Radsim}
\end{figure}

Our first radiation hardness tests of the lanthanum crown glass (NLaK33, S-YGH51 or S-LAH97), used for the middle layer of the early 3-layer lens prototypes, strongly suggested that this material may not be suitable for the final hpDIRC design. Several materials were identified as potential candidates to replace the lanthanum crown glass. One of the leading candidates is lead fluoride PbF$_{2}$ with refractive index 1.78.  

The assessment of the PID performance of DIRC with radiation-hard optical materials is an important aspect of the selection of a suitable lens material. The PbF$_{2}$ sample was visibly less transparent than previously used glass what motivated detailed simulation study of DIRC performance with the lens with PbF$_{2}$ as the middle-layer material. It was implemented in the hpDIRC Geant4 simulation that includes all details about used materials. Figure~\ref{fig:Radsim} shows very promising results of the lanthanum crown glass and PbF$_{2}$ based lens comparison. The evaluated $\pi$/K separation power for 6~GeV/c pions is identical in both cases.

\begin{figure}[htb]
\centerline{%
\includegraphics[width=.95\textwidth]{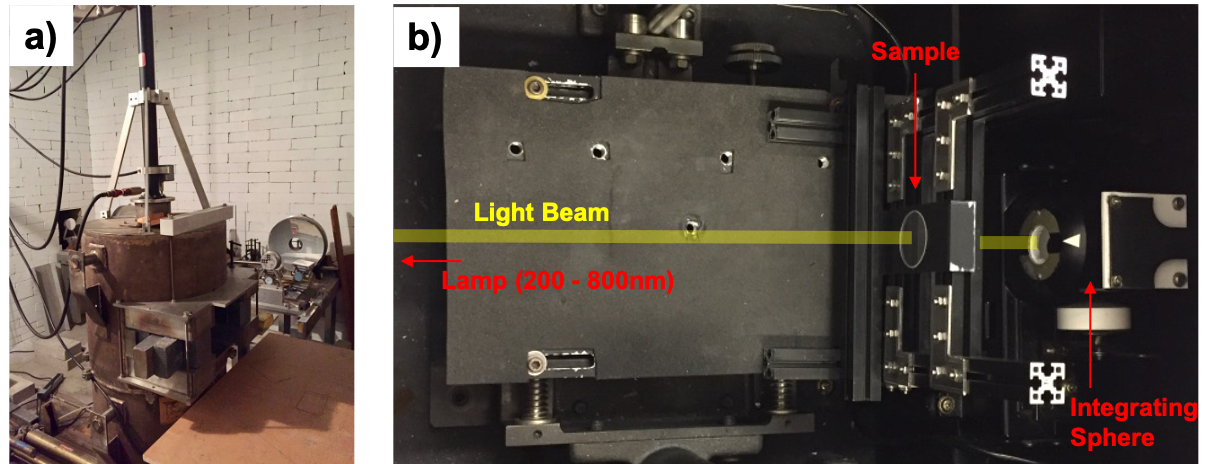}
}%
\caption{Photos of a) the $^{60}$Co source chamber, used to irradiate the optical samples, b) the monochromator setup, used for transmission measurements to quantify radiation damage of the samples. }
\label{fig:RadHard}
\end{figure}

Figure~\ref{fig:RadHard}a shows a dedicated setup for radiation hardness measurements with the $^{60}$Co source, commonly used for studies of radiation hardness of optical materials, prepared at the BNL gamma facility. Several samples, including PbF$_{2}$, were irradiated in a few steps with a calibrated deposited dose of gammas. Figure~\ref{fig:RadHard}b shows the monochromator used to quantify radiation damage. An optical sample is placed in a special fixed holder in front of integrating sphere used to detect light. A set of two lamps is used to generate a light beam in wavelength range 200-800~nm that passes through the optical sample to the integrating sphere. The difference between transmission measurement in the monochromator without any sample, the reference fused silica sample never irradiated, and irradiated samples allows to quantify transmission loss as a function of deposited dose for a wide range of wavelength. Two light sources with different measurement precision are used in the process, resulting in a precision of ±0.5$\%$ for the 390-800±nm range and ±1.2$\%$ below 390~nm.  

\begin{figure}[htb]
\centerline{%
\includegraphics[width=.95\textwidth]{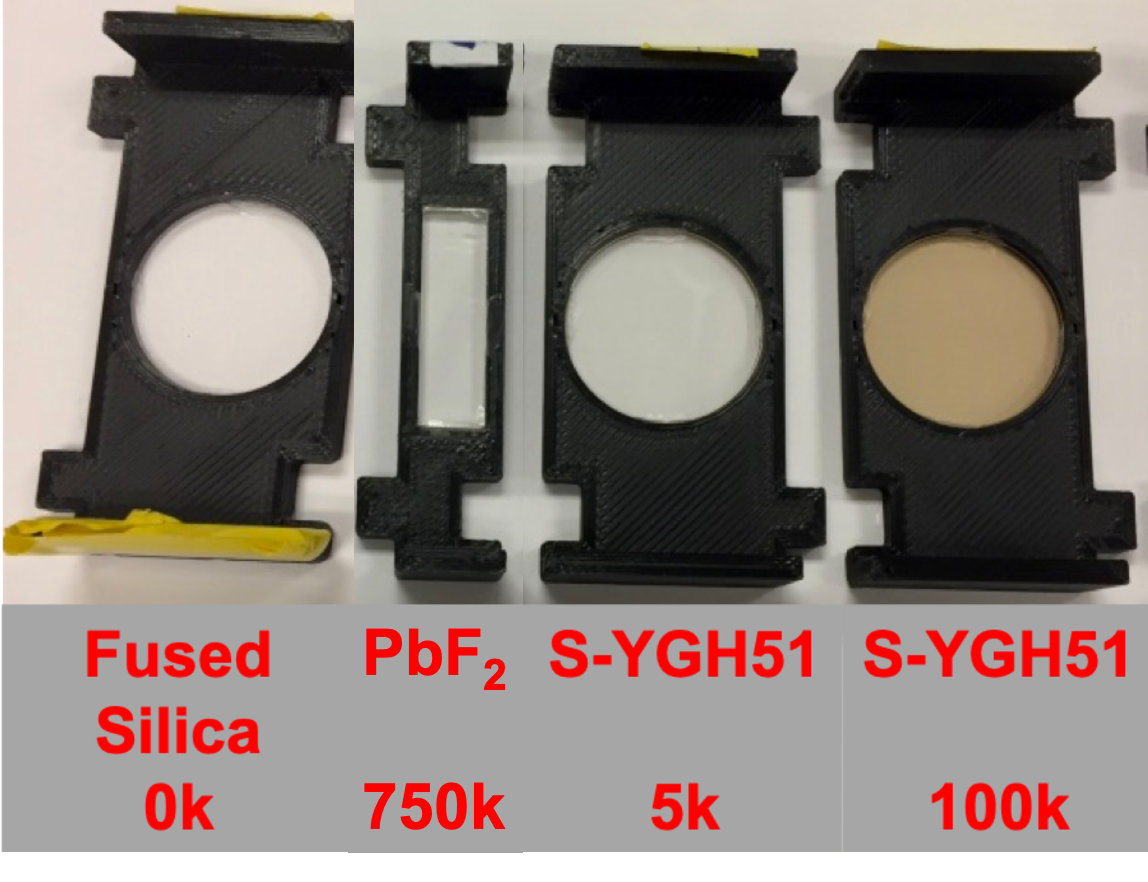}
}%
\caption{Photos of samples of materials after irradiation in a $^{60}$Co source with corresponding deposited doses.
 }
\label{fig:RadHard2}
\end{figure}

Figure~\ref{fig:RadHard2} shows a photo of selected samples tested at BNL after the completion of the first irradiation program, as well as a Fused Silica sample, never irradiated, used as a reference in the transmission measurements. A visible color change was observed only for the S-YGH51 (lanthanum crown glass) sample. The  radiation hardness test of PbF$_{2}$ samples showed very promising results. Currently, a more detailed study of birefringence as well as radio-, UV- and visible luminescence of the candidate materials is being prepared. 

\begin{figure}[htb]
\centerline{%
\includegraphics[width=.95\textwidth]{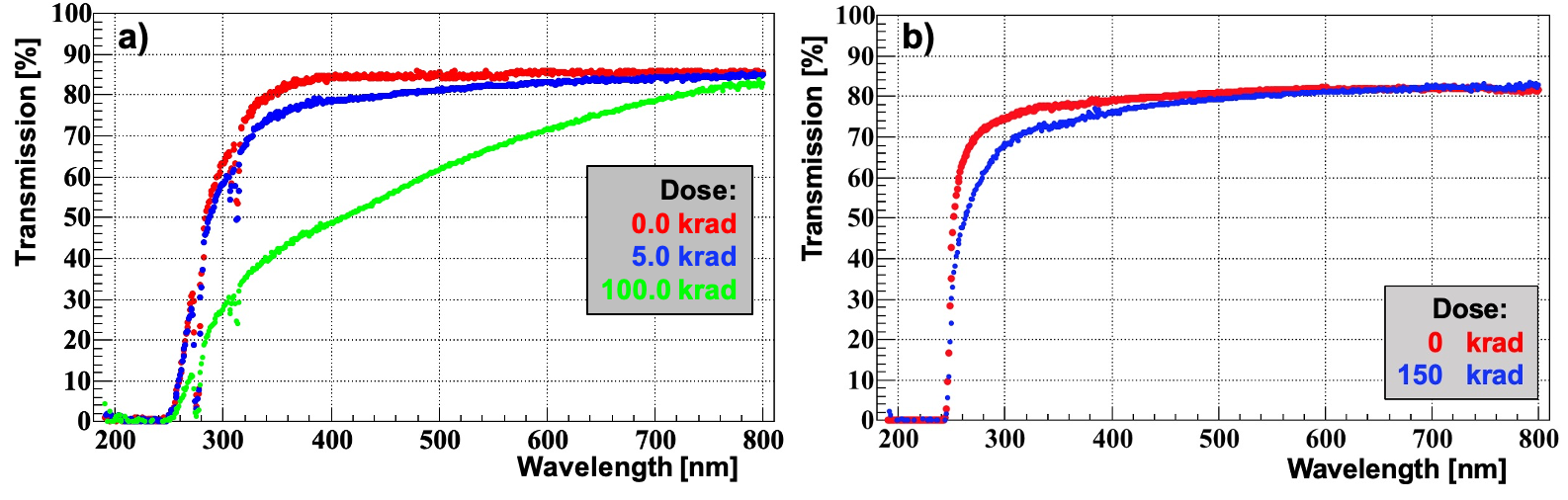}
}%
\caption{Measured transmission (not corrected for Fresnel losses) through the 2~mm-thick S-YGH51 (a) and 4~mm PbF$_{2}$(b) samples as a function of wavelength for selected amounts of deposited gamma ray dose from  $^{60}$Co source.
 }
\label{fig:RadHard3}
\end{figure}

The results for the 2 mm-thick S-YGH51 Lanthanum crown glass sample are shown in Fig.~\ref{fig:RadHard3}a. Smaller irradiation doses for each step were selected based on earlier calibration measurements. We observe a significant deterioration of transmission for the full wavelength range. After 100~krad deposited dose we observe a transmission drop from 85$\%$ to 48$\%$ at 400~nm. The S-YGH51 glass is recovering part of the transmission loss with time. A detailed quantification of the recovery process is in
progress. Figure~\ref{fig:RadHard3}b shows results for the  4~mm-thick PbF$_{2}$ sample. We observe a small transmission loss below 400 nm. The saturation
of the absorption band happened, however, already below the first 50~krad step and we observed no further
changes, even after reaching 150 krad.

The 3-layer lens prototype, with the middle layer made of radiation hard PbF$_{2}$, was designed, purchased, and is being currently produced. This material is very challenging to work with that is why alternative materials are still being studied as producing radiation hard lens is  an essential step for hpDIRC development.

\section{Summary}

The Geant4 simulation studies have shown that the hpDIRC detector is a great match to the PID requirements for the barrel region of the future Electron-Ion Collider central detector. The development of the radiation hard lens, the key technological element for realization of the hpDIRC concept is in progress. Several candidate materials for this lens were identified and a first prototypes are being produced. They will have to be evaluated on test benches. The predicted performance of the hpDIRC detector will have to be validated using a vertical slice prototype in particle beams.

\acknowledgments

This work is supported by EIC R$\&$D BNL under eRD4 and eRD14.


\end{document}